\documentclass[12pt]{iopart}

\usepackage{graphicx}

\newcommand{\cfigl}[3]{\begin{figure}[!hbtp]\centering
 \includegraphics[width=.5\textwidth]{#2}\caption{\small{{#3}}}\label{#1}\end{figure}}
\usepackage{iopams}
\newcommand{\field}[1]{\mathbb{#1}}
\newcommand{\RR}{\field{R}}
\newcommand{\ID}{\field{I}}

\begin{document}

\title[Space-time symmetry group of an elementary particle]{The space-time symmetry group of a \\spin $1/2$ elementary particle}

\author{Mart\'{\i}n Rivas}

\address{Theoretical Physics Department, The University of the Basque Country,\\ 
Apdo.~644, 48080 Bilbao, Spain}
\ead{martin.rivas@ehu.es}

\begin{abstract}
{The space-time symmetry group of a model of a relativistic spin 1/2 
elementary particle, which satisfies Dirac's equation when quantized, is analyzed. 
It is shown that this group, larger than the 
Poincar\'e group, also contains space-time dilations and local rotations. It has two Casimir operators, 
one is the spin and the other is the spin projection on the body frame.  Its 
similarities with the standard model are discussed. If we consider this last spin 
observable as describing isospin, then,
this Dirac particle represents a massive system of spin 1/2 and isospin 
1/2. There are two possible irreducible representations of this kind of particles, 
a colourless or a coloured one, where the colour observable is 
also another spin contribution related to the zitterbewegung. It is the spin, with its twofold structure,
the only intrinsic property of this Dirac elementary particle.}
\end{abstract}


\submitto{\JPA}

\maketitle

\section{Introduction and scope}
The kinematical group of a relativistic formalism is the Poincar\'e group. The Hilbert space
of states of an elementary particle carries an irreducible representation of the Poincar\'e group.
The intrinsic properties of an elementary particle are thus associated to the Casimir 
invariants of its symmetry group. The Poincar\'e group has two Casimir
operators whose eigenvalues define two intrinsic properties
of elementary particles, namely the mass and spin. Elementary matter, in 
addition to mass and spin, has another set of properties
like electric charge, isospin, baryonic and leptonic number, strangeness, etc.,
whose origin is not clearly connected to any space-time symmetry group.
They are supposed to be related to some internal symmetry groups. The adjective
internal is used here to mean that this symmetry 
is coming from some unknown degrees of freedom, whose geometrical interpretation is 
not very clear and perhaps suggesting that they are not connected with any space-time transformations.

In this paper we shall consider a classical Poincar\'e 
invariant model of an elementary spinning particle which satisfies Dirac's equation when quantized.
The aim of this work is to show that it has a larger space-time symmetry group 
than the Poincar\'e group, which will be analyzed in detail, for the classical model and its quantum mechanical
representation.

The kinematical formalism of elementary particles developed by the author
\cite{Rivasbook} defines a classical elementary particle as a 
mechanical system whose kinematical space is a homogeneous space of the kinematical group of 
space-time transformations. The kinematical space of any mechanical 
system is, by definition, the manifold spanned by the variables which define the initial and final state of 
the system in the variational description. 
It is a manifold larger than the configuration space
and it consists of the time variable and the independent degrees of freedom and their time 
derivatives up to one order less
than the highest order they have in the Lagrangian function which defines the action integral. 
Conversely, if the kinematical space
of a mechanical system is known, then, necessarily, the Lagrangian for this system is a 
function of these kinematical variables and their next order time derivative. Please, remark that, as
a general statement, we do not restrict the Lagrangians to depend only on the first time derivatives
of the degrees of freedom. We make emphasis in the description of the classical formalism in terms of the
end point variables of the variational approach.

The advantage of this formalism with respect to other approaches for describing classical 
spinning particles is outlined in chapter 5 of reference \cite{Rivasbook}, 
where a comparative analysis of different models is performed.
One of the features is that any classical system which fulfils with this classical 
definition of elementary particle has the property that, when quantized,
its Hilbert space is a representation space of a projective unitary irreducible representation
of the kinematical group. 
It thus complies with Wigner's definition of an elementary particle 
in the quantum formalism \cite{Rivas2}. 
 
Another is that it is not necessary to postulate any particular Lagrangian. It is sufficient
to fix the kinematical variables (which span a homogeneous space of the Poincar\'e group)
for describing some plausible elementary models. 
In particular, if the variational formalism
is stated in terms of some arbitrary and Poincar\'e invariant evolution parameter $\tau$, then the Lagrangian
of any mechanical system is a homogeneous function of first degree of the $\tau-$derivatives of the 
kinematical variables.
Feynman's quantization shows the importance of the end point variables of the variational formalism.
They are the only variables which survive after every path integral and Feynman's probability
amplitude is an explicit function of all of them and not of the independent degrees of freedom.

The physical interest of this work is to analyze the symmetries of a model of a spinning Dirac particle 
described within this framework which has the property that its center of mass and center of charge are different points.
It has been proven recently that the electromagnetic interaction of two of these particles with the same charge
produces metastable bound states provided the spins of both particles are parallel and their relative
separation, center of mass velocity and internal phase fulfill certain requirements \cite{dyn}. Electromagnetism does not
allow the possibility of formation of bound pairs for spinless point particles of the same charge. Nevertheless,
two of these spinning particles, although the force between the charges is repulsive, it can become
atractive for the motion of their center of masses, provided the two 
particles are separated below Compton's wavelength.
Electromagnetism does not forbid the existence of bound pairs of this kind of spinning Dirac particles.

Because the variational formalism when written in terms of the kinematical variables 
seems to be not widespread we shall briefly
describe some of its basic features in sections 2 to 4. Very technical details are left for a 
reading of the original publications.
Section 2 is devoted to a particular parameterization
of the Poincar\'e group in order to properly interpret the geometrical meaning of the variables which
span the different homogeneous spaces. These variables will be treated as the kinematical variables
of the corresponding elementary systems. In section 3 we consider the classical model which satisfies
Dirac's equation when quantized and we describe its main features. This Dirac particle is 
a system of six degrees of freedom.
Three represent the position of a point (the position of the charge of the particle) 
which is moving at the speed of light and the other three represent
its orientation in space. It is a point-like particle with orientation. 
The Lagrangian will also depend on the acceleration of the point
and of the angular velocity of its local oriented frame.
Section 4 shows how Feynman's quantization
of the kinematical formalism describe the wave function and the differential structure of the generators
of the symmetry groups in this representation.

In section 5 we analyze the additional space-time symmetries of the model, which reduce to the space-time
dilations and the local rotations of the body frame. Section 6 deals with the analysis of the complete
space-time symmetry group, its Casimir operators and the irreducible representations for this spin $1/2$
Dirac particle.
Once the symmetry group has been enlarged, a greater kinematical space can be defined. Chapter 7 considers
the requirement of enlarging the kinematical space, by adding a classical phase, in order for the system
to still satisfy Dirac's equation.
Finally, section 8 contains a summary of the main results and some final comments about the relationship
between this new symmetry group and the standard model.
The explicit calculation of the structure of the generators is included in the Appendix.

\section{The kinematical variables}

Any group element $g$ of the Poincar\'e group can be parameterized 
by the ten variables $g\equiv(t,{\bi r},{\bi v},\balpha)$
which have the following dimensions and domains: $t\in\RR$ is a time variable 
which describes the time translation,
${\bi r}\in \RR^3$ are three position variables associated to the three-dimensional space
translations, ${\bi v}\in\RR^3$, with $v<c$, are three velocity parameters which define the relative
velocity between observers and, finally, the three dimensionless $\balpha\in SO(3)$, which 
represent the relative orientation between the corresponding Cartesian frames. These dimensions
will be shared by the variables which define the corresponding 
homogeneous spaces of the Poincar\'e group. 

The manifold spanned by the variables
$t$ and ${\bi r}$ is the space-time manifold, which is clearly a homogeneous space of the Poincar\'e
group. According to our definition of elementary particle, 
it represents the kinematical space of the spinless point particle. These variables
are interpreted as the time and position of the particle. The corresponding Lagrangian will be, in general,
a function of $t$, ${\bi r}$ and its time derivative, the velocity of the point ${\bi r}$. 
It can be easily deduced
from invariance requirements and it is uniquely defined up to a total $\tau-$derivative \cite{Rivasbook}.

To describe spinning particles we have to consider larger homogeneous spaces 
than the space-time manifold. The Poincar\'e group has three maximal homogeneous
spaces spanned by the above ten variables but with some minor restrictions. 
One is the complete group manifold, where the parameter $v<c$. 
Another is that manifold with the constraint $v=c$, and it will therefore describe
particles where the point ${\bi r}$ is moving at the speed of light, 
and finally the manifold where $v>c$ and this allows to describe
tachyonic matter. The remaining variables $t$, ${\bi r}$ and $\balpha$ have the same domains as above.
In all three cases, the initial and final state of a classical elementary particle in the variational approach
is described by the measurement of a time $t$, 
the position of a point ${\bi r}$, the velocity of this point ${\bi v}$ and the orientation
of the particle $\balpha$, which can be interpreted as 
the orientation of a local instantaneous
frame of unit axis ${\bi e}_i$, $i=1,2,3$, with origin at the point ${\bi r}$. 
The difference between the three
cases lies in the intrinsic character of the velocity parameter $v$ which can be used to 
describe particles whose position ${\bi r}$ moves always 
with velocity below, above or equal to $c$. 
The Lagrangian for describing these systems will be in general, a function
of these ten variables, with the corresponding constraint on the $v$ variable, 
and also of their time derivatives. It will be thus also a function of the acceleration
of the point ${\bi r}$ and of the angular velocity $\bomega$ of the motion of the body local frame.

This is what the general kinematical formalism establishes. The systems which satisfy Dirac's equation when
quantized, correspond to the systems for which $v=c$. For these systems the point ${\bi r}$ represents the
position of the charge of the particle, but not the centre of mass ${\bi q}$, which, for a spinning particle, 
results a different point than ${\bi r}$. This separation is analytically related to the dependence of the Lagrangian on the
acceleration. The spin structure is related to this separation between
${\bi r}$ and ${\bi q}$ and its relative motion and, 
also to the rotation of the body frame with angular velocity
$\bomega$. This separation between the center of mass and center of charge of a spinning particle
is a feature also shared by some other spinning models which can be found in 
chapter 5 of reference \cite{Rivasbook}.

All these particle models represent point-like objects 
with orientation but whose centre of mass
is a different point than the point where the charge of the particle is located. 
The classical dynamical equation satisfied by the position ${\bi r}$ 
of these systems was analyzed in \cite{dyn}.
The quantization of the models with $v=c=1$ shows that the corresponding 
quantum systems are only spin $1/2$ particles.
No higher spin elementary systems are obtained when using that manifold as a kinematical space. 
In the tachyonic case only spin 1 is allowed. For particles with $v<c$ the maximum spin is only $3/2$.

\section{The classical model}

Let us consider any regular Lagrangian system whose Lagrangian depends on time $t$, on the $n$ independent degrees
of freedom $q_i$, $i=1,\ldots,n$ and their time derivatives up to a finite order $k$, $q_i^{(k)}\equiv d^kq_i/dt^k$. We 
denote the kinematical variables in the following form:
\[
x_0=t,\quad x_i=q_{i},\quad x_{n+i}=q^{(1)}_i,\quad\ldots\quad x_{(k-1)n+i}=q^{(k-1)}_i,\quad i=1,\ldots,n.
\]
This system of $n$ degrees of freedom has a kinematical space of dimension $kn+1$.
We can always write this Lagrangian in terms of the kinematical variables $x$ 
and their next order $\tau-$derivative $\dot{x}$, 
where $\tau$ is any arbitrary, Poincar\'e invariant, evolution parameter. 
The advantage of working with an arbitary evolution parameter is that
the Lagrangian expressed in this way becomes a homogeneous function of first degree 
of the $\tau-$derivatives of the kinematical variables. In fact, each time derivative $q_i^{(r)}$ can be written as a quotient
of two $\tau-$derivatives of two kinematical variables $q_i^{(r)}=\dot{q}_i^{(r-1)}/\dot{t}$, and thus
\[
\int_{t_1}^{t_2}L(t,q,\ldots,q^{(k)})dt=\int_{\tau_1}^{\tau_2}L(t,q,\dot{q}/\dot{t},\ldots,\dot{q}^{(k-1)}/\dot{t})\dot{t}d\tau=
\int_{\tau_1}^{\tau_2} L(x,\dot{x})d\tau.
\]
It is the last $\dot{t}$ term which makes the integrand a homogeneous function of first degree of the derivatives
of the kinematical variables. 
It thus satisfies Euler's theorem for homogeneous functions
 \begin{equation}
L(x,\dot{x})=\frac{\partial L}{\partial\dot{x}_i}\dot{x}_i\equiv F_i(x,\dot{x})\dot{x}_i,\quad i=0,1,\ldots,kn.
 \label{eq:hom}
 \end{equation}
The generalized coordinates
are the degrees of freedom and their time derivatives up to order $k-1$, so that there are up to $kn$ generalized
momenta $p_{is}$ which are defined by
\[
p_{is}=\sum_{r=0}^{k-s}\frac{d^r}{dt^r}F_{(r+s-1)n+i},\quad i=1,\ldots,n,\quad s=1,\ldots,k.
\]
The Hamiltonian is
\[
H=p_{is}q_{i}^{(s)}-L,
\]
and the phase space is thus of dimension $2kn$.

If $G$ is a $r-$parameter symmetry group of parameters $g^\alpha$, $\alpha=1,\ldots,r$, 
which leaves invariant the Lagrangian and transforms
infinitesimally the kinematical variables in the form
\[
x'_j=x_j+M_{j\alpha}(x)\delta g^\alpha,\quad j=0,1,\ldots, kn,\quad \alpha=1,\ldots,r,
\]
the $r$ conserved Noether's observables are given by
\[
N_\alpha= HM_{0\alpha}(x)-p_{is}M_{\{(s-1)n+i\}\alpha}(x),\quad i=1,\ldots,n,\quad s=1,\ldots k.
\]
The advantage of this formulation is that we can obtain general expressions for the conserved quantities
in terms of the above $F_i(x,\dot{x})$ functions, and their time derivatives, which are homogeneous functions of zero-th degree
of the variables $\dot{x}_i$ and of the way the kinematical variables transform $M_{j\alpha}(x)$.

Let us consider now any Lagrangian system whose kinematical space is spanned by the variables $(t,{\bi r},{\bi v},\balpha)$, with 
$v=c$. The general form of any Lagrangian of this kind of systems is
\[
L=\dot{t}T+{\dot{\bi r}}\cdot{\bi R}+{\dot{\bi v}}\cdot{\bi V}+{\dot{\balpha}}\cdot{\bi A},
\]
where according to the above homogeneity property (\ref{eq:hom}), $T={\partial L}/{\partial\dot{t}}$, ${\bi R}={\partial L}/{\partial\dot{\bi r}}$,
${\bi V}={\partial L}/{\partial\dot{\bi v}}$ and ${\bi A}={\partial L}/{\partial\dot{\balpha}}$.
If instead of ${\dot{\balpha}}$ we consider the angular velocity $\bomega$, which is a linear function of the ${\dot{\balpha}}$,
the last term ${\dot{\balpha}}\cdot{\bi A}$, can be transformed into
$\bomega\cdot{\bi W}$, where $W_i={\partial L}/{\partial \omega_i}$.

We thus get, the following conserved quantities from the invariance
of the Lagrangian under the Poincar\'e group. These expressions are independent of the particular Lagrangian we take, as far
as we keep fixed the corresponding kinematical space. 
The energy and linear momentum
 \begin{equation}
H=-T-{\bi v}\cdot\,\frac{d{\bi V}}{dt},\quad {\bi P}={\bi R}-\frac{d{\bi V}}{dt},
 \label{eq:HP}
 \end{equation}
and the kinematical and angular momentum are, respectively
 \begin{equation}
{\bi K}=H{\bi r}-{\bi P}t+{\bi v}\times{\bi S},\quad {\bi J}={\bi r}\times{\bi P}+{\bi S},\quad {\bi S}={\bi v}\times{\bi V}+{\bi W}.
 \label{eq:KJ}
 \end{equation}
${\bi S}$ is the translation invariant part of the angular momentum. It is the classical equivalent
of Dirac's spin observable. It is the sum of two parts ${\bi S}={\bi Z}+{\bi W}$. The ${\bi Z}={\bi v}\times{\bi V}$ part
comes from the dependence of the Lagrangian on $\dot{\bi v}$ and the ${\bi W}$ part from the dependence on the angular
velocity.
The time derivative of the conserved ${\bi J}$ leads to $d{\bi S}/dt={\bi P}\times{\bi v}$.
Since in general the structure of the linear momentum, given in (\ref{eq:HP}), shows 
that ${\bi P}$ and ${\bi v}$ are not collinear vectors, the spin ${\bi S}$ is only constant
in the centre of mass frame. In this frame ${\bi S}$ is a constant vector and $H=\pm m$, ${\bi P}={\bi K}=0$, 
so that for positive energy $H=m$ particles we get, 
from the expression of the kinematical momentum (\ref{eq:KJ}), the dynamical equation for the point ${\bi r}$
 \begin{equation}
{\bi r}=\frac{1}{m}{\bi S}\times{\bi v}.
 \label{eq:dynam}
 \end{equation}
The point ${\bi r}$ moves in circles at the speed of light on a plane orthogonal to the constant spin, 
as depicted in figure \ref{fig:elec}. For negative energy $H=-m$ we get the time reversed motion.

\cfigl{fig:elec}{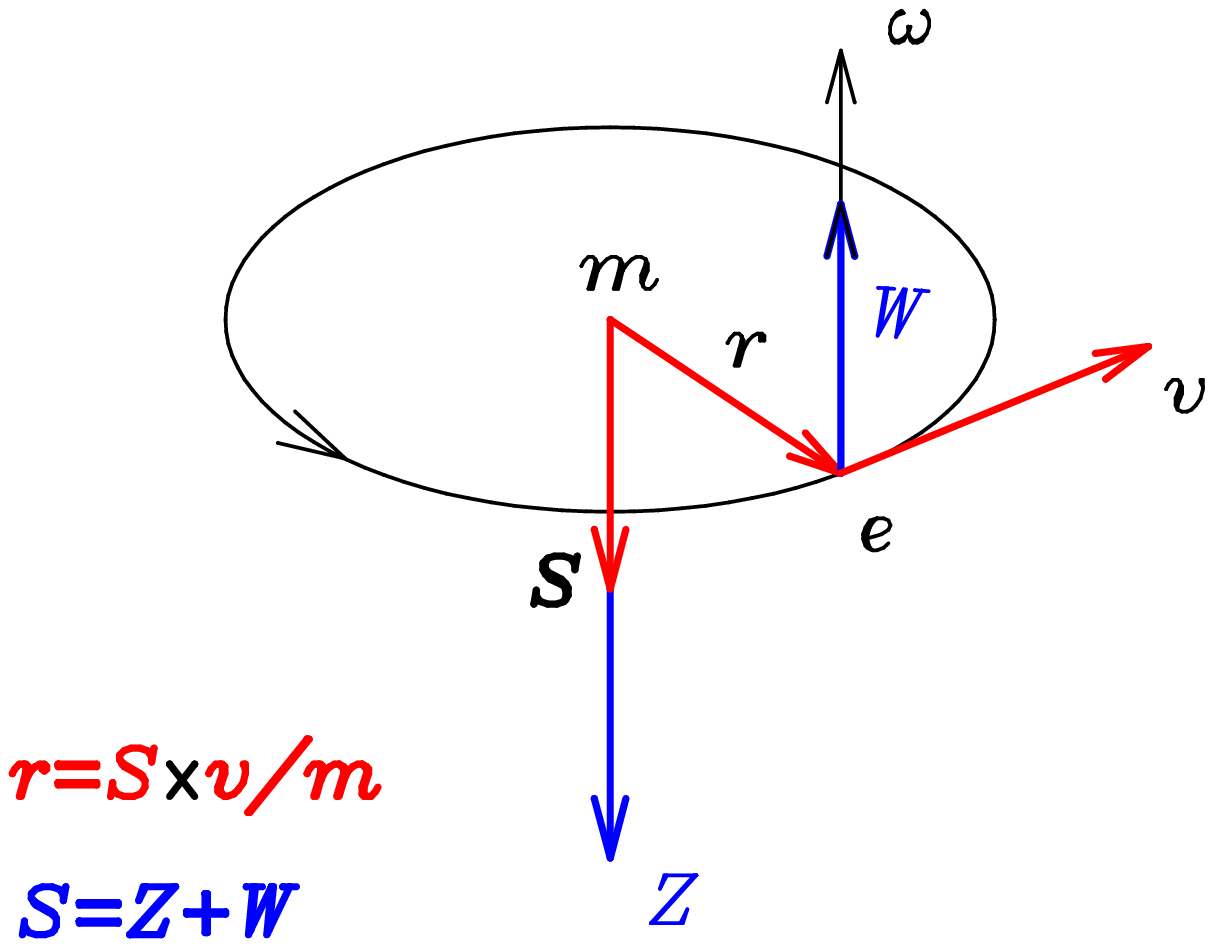}{Motion of the charge of the particle ($H=m>0$), in the centre of mass frame. 
The total spin ${\bi S}$ is the sum of the orbital part ${\bi Z}$ and the rotational part of the body frame ${\bi W}$.
It is not depicted the local body frame, with origin at point ${\bi r}$, ${\bi e}_i$, $i=1,2,3$,
which rotates with angular velocity $\bomega$. The motion of the charge, with respect to the fixed spin
direction, is left-handed.}

The motion of the charge around the
centre of mass is a circular motion, known as the zitterbewegung, of radius $R=S/mc$, {\it i.e.}, 
half Compton's wavelength when quantized, 
and angular frequency $\omega=mc^2/S$.
Since the energy $H$ is not definite positive we can describe matter and antimatter. 
They have a different chirality. Matter is left-handed while antimatter is right-handed.
For matter, once the spin direction is fixed, the motion of the charge is counterclockwise
when looking along the spin direction. The phase of this internal motion of the charge
is increasing in the opposite direction to the usual sign convention for $p$-forms. 
The motion is clockwise for antimatter.  

The classical expression equivalent to Dirac's equation is obtained by taking the 
time derivative of the kinematical momentum ${\bi K}$ in (\ref{eq:KJ}), 
and a subsequent scalar product with ${\bi v}$, {\it i.e.}, the linear relationship between
$H$ and ${\bi P}$
 \begin{equation}
H={\bi P}\cdot{\bi v}+{\bi S}\cdot\left(\frac{d{\bi v}}{dt}\times{\bi v}\right).
 \label{eq:Dir}
 \end{equation}
It is the quantum representation of this relationship between observables when acting on the wavefunction of the
system, which produces Dirac's equation \cite{Rivas2}.

\section{Quantization of the model}

If we analyze this classical particle in the centre of mass frame it becomes a mechanical 
system of three degrees of freedom.
These are the $x$ and $y$ coordinates of the point charge on the plane and the phase $\alpha$
of the rotation of the body axis with angular velocity $\omega$. But this phase is the same as the phase of the
orbital motion and because this motion is a circle of constant radius only one degree of freedom is left,
for instance, the $x$ coordinate.
In the centre of mass frame the system is thus equivalent to a one-dimensional harmonic oscillator 
of angular frequency $\omega=mc^2/S$ in its ground state. The ground energy of this one-dimensional
harmonic oscillator $\hbar\omega/2=mc^2$ for particles, so that
the classical constant parameter $S=\hbar/2$. All Lagrangian systems defined with this kinematical space
have this behaviour and represent spin $1/2$ particles when quantized.
If this model represents an elementary particle it has no excited states and thus no higher spin model
can be obtained which has the same kinematical space as this one.

When quantizing any mechanical system described by means of this kinematical formalism, through Feynman's path integral 
approach \cite{Rivas2}, the quantization
leads to the following results:
\begin{enumerate}
\item{If $x$ are the kinematical variables of the variational approach, 
Feynman's kernel $K(x_1,x_2)$ which describes the probability amplitude for the evolution of the system
between the initial point $x_1$ to the final point $x_2$, is only a function (more properly a distribution)
of the end point kinematical variables.}
\item{ The wave function of the quantized system
is a complex squared integrable function of these variables $\psi(x)$, with respect to some suitable invariant measure
over the kinematical space.}
\item{If $G$ is a symmetry group of parameters $g^\alpha$, which transform infinitesimally 
the kinematical variables in the form:
\[
x'_j=x_j+M_{j\alpha}(x)\delta g^\alpha,
\]
the representation of the generators is given by the self-adjoint operators
\[
X_\alpha=-iM_{j\alpha}(x)\frac{\partial }{\partial x_j}.
\]}
\end{enumerate}
In our model of the Dirac particle, the wave function becomes a complex squared integrable function 
defined on the kinematical space $\psi(t,{\bi r},{\bi v},\balpha)$.
The Poincar\'e group unitary realization over the corresponding Hilbert space 
has the usual selfadjoint generators. They are represented by the differential operators,
with respect to the kinematical variables,
which are obtained in detail in the Appendix:
\[
H=i{\partial }/{\partial t},\quad P_i=-i{\partial }/{\partial r_i},\]
\[
K_i=ir_i{\partial }/{\partial t}+it{\partial }/{\partial r_i}+\epsilon_{ijk}v_jS_k,
\quad J_i=-i\epsilon_{ijk}r_j{\partial }/{\partial r_k}+S_j
\]
or in three-vector form
\[
H=i{\partial }/{\partial t},\quad {\bi P}=-i\nabla,\]
\[
 {\bi K}=i{\bi r}{\partial }/{\partial t}+it{\nabla}+{\bi v}\times{\bi S},\quad {\bi J}=-i{\bi r}\times\nabla+{\bi S}={\bi L}+{\bi S}.
\]
The spin operator ${\bi S}$ is given by
\[
S_i=-i\epsilon_{ijk}v_j{\partial }/{\partial v_k}+W_i,\quad\hbox{\rm or}\quad{\bi S}=-i{\bi v}\times{\nabla_v}+{\bi W}={\bi Z}+{\bi W}.
\]
$\nabla_v$ is the gradient operator with respect to the $v_i$ variables and the ${\bi W}$ operator involves 
differential operators with respect to the orientation variables. Its structure depends 
on the election of the variables which represent the orientation and which correspond 
to the different parameterizations
of the rotation group. In the normal or canonical parameterization of the rotation group, every rotation
is characterized by a three vector $\balpha=\alpha{\bi n}$, where ${\bi n}$ is a unit vector along the rotation axis
and $\alpha$ the clokwise rotated angle. If we represent the unit vector ${\bi n}$ by the usual polar 
and azimuthal angles $(\theta,\phi)$,
$\theta\in[0,\pi]$ and $\phi\in[0,2\pi]$,
then every rotation is parameterized by the three dimensionless variables $(\alpha,\theta,\phi)$. 
When acting the rotation group on this manifold, the ${\bi W}$ operators take the form given in
the Appendix (\ref{eq:Y1})-(\ref{eq:Y3}).

With the definitions $p^\mu\equiv(H,{\bi P})$ and $J^{0i}=-J^{i0}=K_i$, $J^{ij}=-J^{ji}=\epsilon^{ijk}J_k$,
\[
[p^\mu,p^\nu]=0,\quad [J^{\mu\nu},J^{\sigma\rho}]=i\eta^{\mu\sigma}J^{\nu\rho}+i\eta^{\nu\rho}J^{\mu\sigma}
-i\eta^{\mu\sigma}J^{\nu\rho}-i\eta^{\nu\rho}J^{\mu\sigma},\]
\[
 [J^{\mu\nu},p^{\rho}]=i\eta^{\mu\rho}p^{\nu}-i\eta^{\nu\rho}p^\mu,
\]
where $\eta^{\mu\nu}$ is Minkowski's metric tensor, are the usual commutation relations 
of the Poincar\'e group.

The angular momentum operator ${\bi J}$ contains the orbital angular momentum operator ${\bi L}$ 
and the spin part ${\bi S}$, which is translation invariant and which has a twofold structure. One, ${\bi Z}$, has the form
of an orbital angular momentum $-i{\bi v}\times\nabla_v$ in terms of the velocity variables. 
It is related to the zitterbewegung part of the spin and quantizes with integer eigenvalues. Finally, 
another related to the orientation variables ${\bi W}$ which can have either integer and half integer
eigenvalues and which is related, in the classical case, to the rotation of the particle.
They satisfy the commutation relations
\[
[Z_i,Z_j]=i\epsilon_{ijk}Z_k,\quad [W_i,W_j]=i\epsilon_{ijk}W_k,\quad [Z_i,W_k]=0,\quad [S_i,S_j]=i\epsilon_{ijk}S_k.
\]
The structure of the generators of rotations ${\bi J}$ contains differential operators with respect to the kinematical 
variables which are transformed when we rotate observer's axis, so that the ${\bi L}$ part is associated to the change
of the variable ${\bi r}$, the ${\bi Z}$ part is associated to the change of the velocity ${\bi v}$
under rotations and finally the ${\bi W}$ part contains the contribution of the change of the orientation of the body
frame.

\section{Additional space-time symmetries}

Up to this point we have outlined the main classical and quantum mechanical features of the 
kinematical formalism and of the model of the elementary spinning particle of spin 1/2 
we want to further analyze.

The kinematical variables of this classical spinning elementary particle 
are reduced to time $t$, position
${\bi r}$, velocity ${\bi v}$ and orientation $\balpha$, but the velocity is always $v=c$. 
It is always $1$ in natural units. If the particle has
mass $m\neq0$ and spin $s\neq0$, we can also define a natural unit of length $s/mc$ and a natural unit of time $s/mc^2$.
The unit of length is the radius of the zitterbewegung motion of figure \ref{fig:elec}, and the unit of time 
is the time employed by the charge, in the centre of mass frame, during a complete turn. 
This implies that the
whole set of kinematical variables and their time derivatives 
can be taken dimensionless, and the classical formalism is therefore 
invariant under space-time dilations
which do not modify the speed of light.

It turns out that although we started with the Poincar\'e group as the basic 
space-time symmetry group, this kind of massive spinning Dirac particles,
has a larger symmetry group. It also contains at least 
space-time dilations with generator $D$.
The new conserved Noether observable takes the form
 \begin{equation}
D=tH-{\bi r}\cdot{\bi P}.
 \label{eq:DD}
 \end{equation}

Let $R(\bbeta)$ be an arbitrary rotation which changes observer's axes.
The action of this arbitray rotation $R(\bbeta)$ on the kinematical variables is
\[
t'=t,\quad {\bi r}'=R(\bbeta){\bi r},\quad {\bi v}'=R(\bbeta){\bi v},\quad R(\alpha')=R(\bbeta)R(\balpha),
\]
and this is the reason why the generators ${\bi J}$ of rotations involve differential operators with respect to 
all these variables, the time excluded.

The orientation of the particle, represented by the variables $\balpha$, or the equivalent orthogonal
rotation matrix $R(\balpha)$, is interpreted as the orientation of an hypothetical Cartesian
frame of unit axis ${\bi e}_i$, $i=1,2,3$, located at point ${\bi r}$. 
It has no physical reality but can be interpreted as the corresponding
Cartesian frame of some instantaneous inertial observer with origin at that point. 
But the election 
of this frame is completely arbitrary so that the formalism has to be independent of its actual value.
This means that, in addition to the above rotation group which modifies the laboratory axes, there will
be another rotation group of elements $R(\bgamma)$ 
which modifies only the orientation variables $\balpha$, without modifying the variables
${\bi r}$ and ${\bi v}$, {\it i.e.}, the rotation only of the body frame: 
 \begin{equation}
t'=t,\quad {\bi r}'={\bi r},\quad {\bi v}'={\bi v},\quad R(\alpha')=R(\bgamma)R(\balpha),
 \label{eq:roto}
 \end{equation}
The generators of this new rotation group, 
which affects only to the orientation variables, will be the projection of the angular momentum generators 
${\bi W}$ onto the body axes. From Noether's theorem the corresponding classical conserved observables
are
 \begin{equation}
T_i={\bi W}\cdot{\bi e}_i,
 \label{eq:isos}
 \end{equation}
where the ${\bi e}_i$ are the three orthogonal unit vectors which define the body axis.

If $R(\balpha)$ is the orthogonal rotation matrix which describes the orientation of the particle,
when considered by columns
these columns describe the components of the three orthogonal unit vectors ${\bi e}_i$, $i=1,2,3$.
The equations (\ref{eq:roto}) correspond to the transformation ${\bi e}'_i=R(\bgamma){\bi e}_i$
of the body frame.

The $W_i$ operators 
represent the components of the angular momentum operators associated to the change of orientation 
of the particle and projected in the laboratory frame.
The corresponding conserved quantities (\ref{eq:isos}) 
are the components of the angular momentum operators projected 
onto the body frame $T_i={\bi e}_i\cdot{\bi W}$.
When quantizing the system they are given by the differential 
operators (\ref{eq:T1})-(\ref{eq:T3}) of the Appendix
and satisfy
\[
T^2=W^2,\quad [T_i,T_j]=i\epsilon_{ijk}T_k,\]
\[
 [T_i,K_j]=[T_i,J_j]=[T_i,H]=[T_i,D]=[T_i,P_j]=0.
\]
We can see that the selfadjoint operators $T_i$ generate another $SU(2)$ group which is the representation
of the rotation group which modifies only the orientation variables,
commutes with the rotation group generated by the $J_j$, and
with the whole enlarged Poincar\'e group, including space-time dilations. 

Since we expect that the formalism is independent 
of the orientation variables we have another $SO(3)$ group of space-time symmetries of the particle.

\section{Analysis of the enlarged symmetry group}

Let $H$, ${\bi P}$, ${\bi K}$ and ${\bi J}$ be the generators of the Poincar\'e group ${\cal P}$.
With the usual identification of $p^\mu\equiv(H,{\bi P})$ as the four-momentum operators and 
$w^\mu\equiv({\bi P}\cdot{\bi J},H{\bi J}-{\bi K}\times{\bi P})$ as the Pauli-Lubanski four-vector 
operator, the two Casimir operators of the Poincar\'e group are
\[
C_1=p_\mu p^\mu,\quad C_2=-w_\mu w^\mu.
\]
These two Casimir operators, if measured in the centre of mass frame where ${\bi P}={\bi K}=0$, reduce respectively
in an irreducible representation to $C_1=m^2$ and $C_2=H^2J^2=m^2 s(s+1)$.
The two parameters $m$ and $s$, which characterize every irreducible representation of the Poincar\'e group, represent
the intrinsic properties of a Poincar\'e invariant elementary particle.

Let us consider the additional space-time dilations of generator $D$. 
The action of this transformation on the kinematical variables is
\[
t'=e^\lambda t,\quad {\bi r}'=e^\lambda {\bi r},\quad {\bi v}'={\bi v},\quad \balpha'=\balpha.
\]
Let us denote this enlargement of the Poincar\'e group, sometimes called the Weyl group, by ${\cal P}_D$. 
In the quantum representation, this new generator
when acting on the above wavefunctions, has the form:
 \begin{equation}
D=it{\partial }/{\partial t}+i{\bi r}\cdot\nabla.
 \label{eq:D}
 \end{equation}
It satisfies
\[
[D,p^{\mu}]=-i p^\mu,\quad [D,J^{\mu\nu}]=0.
\]
This enlarged group has only one Casimir operator (see \cite{Abellanas}) which, for massive systems where the operator
$C_1\neq 0$ is invertible, it is reduced to 
\[
C=C_2C_1^{-1}=C_1^{-1}C_2\equiv C_2/C_1=s(s+1).
\]
In the centre of mass frame this operator is reduced to $C=S^2$, the squared of the spin operator.

By assuming also space-time dilation invariance this implies that the mass is not an intrinsic property. 
It is the spin the only intrinsic property of this elementary particle. In fact, since the radius
of the internal motion is $R=s/mc$, a change of length and time scale corresponds to a change of 
mass while keeping $s$ and $c$ constants. By this transformation the elementary particle of spin $1/2$ 
modifies its internal radius and therefore its mass
and goes into another mass state.

The structure of the differential operator ${\bi J}={\bi r}\times{\bi P}+{\bi Z}+{\bi W}$, where the spin part
${\bi S}={\bi Z}+{\bi W}$ has only $s=1/2$ eigenvalue for the above model, implies that the eigenvalue of the
$W^2$ corresponds to $w=1/2$ while for the ${\bi Z}^2$ part can be reduced to the two 
possibilities $z=0$ or $z=1$.

In addition to the group ${\cal P}_D$ we also consider the representation of the 
local rotation group generated by the $T_i$
with eigenvalue $t=w=1/2$. We have thus a larger space-time symmetry group with 
an additional $SU(2)$ structure when quantized.

The generators $T_i$ commute with all generators of the group ${\cal P}_D$, 
and this new symmetry group can be written as ${\cal P}_D\otimes SU(2)_T$.

This new group has only two Casimir operators $S^2$ and $T^2$ of eigenvalues $1/2$. This justifies 
that our wavefunction will be written as a four-component wavefunction. When choosing 
the complete commuting set
of operators to classify its states we take the operator $T^2=S^2$, the $S_3$ and 
$T_3$ which can take the values $\pm 1/2$ 
and for instance the $p^\mu p_\mu$ and the $p^\mu$. In this way we can separate in the wavefunction 
the orientation and velocity variables from the space-time variables,
\[
\psi(t,{\bi r},{\bi v},{\balpha})=\sum_{i=1}^{i=4}\phi_i(t,{\bi r})\chi_i({\bi v},{\balpha})
\]
where the four $\chi_i({\bi v},{\balpha})$ can be classified according to the eigenvalues $|s_3,t_3>$.
The functions $\phi_i(t,{\bi r})$ can be chosen as eigenfunctions of the Klein-Gordon operator \cite{Rivasbook}
\[
p_\mu p^\mu \phi_i(t,{\bi r})=m_i^2\phi_i(t,{\bi r}).
\]
Because this operator $p_\mu p^\mu$ does not commute with the $D$ observable, 
the mass eigenvalue $m_i$ is not
an intrinsic property and the corresponding value depends on the particular state $\phi_i$ we consider.

For the classification of the $\chi_i({\bi v},{\balpha})$ states we have also to consider the ${\bi Z}$
angular momentum operators. Because $[Z^2,S^2]=[Z^2,T^2]=[Z^2,p^\mu]=0$, 
we can choose $Z^2$ as an additional commuting observable. It can only take integer eigenvalues
when acting on functions of the velocity variables, because it has the structure of
an orbital angular momentum.
But because the total spin ${\bi S}={\bi Z}+{\bi W}$,
and the $S^2$ has eigenvalue $1/2$, the possible eigenvalues of $Z^2$ can be $z=0$ or $z=1$. See the Appendix
for the possible clasification of the $\chi_i({\bi v},\balpha)$ part, according to $z=0$ which gives rise to the
(\ref{eq:Fi1}-\ref{eq:Fi4}) eigenfunctions, and the $z=1$ eigenfunctions (\ref{eq:Fit1}-\ref{eq:Fit4}). 
In this last case
the eigenfunctions cannot be simultaneously eigenfunctions of $Z_3$. Nevertheless the expectation value of $Z_3$
in the $z=0$ basis vectors $\Phi_i$ is 0, while its expectation value in the $z=1$ basis $\Psi_i$
is $\pm2/3$. 

\section{Enlargement of the kinematical space}

Once the kinematical group has been enlarged by including space-time dilations, we have a new dimensionless
group parameter asociated to this one-parameter subgroup which can also be
used as a new kinematical variable, to produce
a larger homogeneous space of the group. In fact, if we take 
the time derivative of the constant of the motion (\ref{eq:DD}) we get
\[
H={\bi P}\cdot{\bi v}.
\]
If we compare this with the equation (\ref{eq:Dir}), one term is lacking. 
This implies that we need, from the classical point of view, an additional
kinematical variable, a dimensionless phase $\beta$, such that under the action of this new transformation
the enlarged kinematical variables transform
\[
t'=e^\lambda t,\quad {\bi r}'=e^\lambda {\bi r},\quad {\bi v}'={\bi v},\quad \balpha'=\balpha,\quad \beta'=\lambda+\beta.
\]
From the group theoretical point of view this new dimensionless variable corresponds to the normal dimensionles 
group parameter of the transformation generated by $D$.

From the Lagrangian point of view, the new Lagrangian has also to depend on $\beta$ and $\dot{\beta}$, with a general structure
\[
L=\dot{t}T+{\dot{\bi r}}\cdot{\bi R}+{\dot{\bi v}}\cdot{\bi V}+{\bomega}\cdot{\bi W}+\dot{\beta}B,
\]
with $B={\partial L}/{\partial\dot{\beta}}$. The constant of the motion associated to the invariance of the dynamical equations
under this new transformation implies that
\[
D=tH-{\bi r}\cdot{\bi P}-B,
\]
and the new generator in the quantum version takes the form
\[
D=it{\partial }/{\partial t}+i{\bi r}\cdot\nabla+i\frac{\partial }{\partial\beta}.
\]
In this way the last term of (\ref{eq:Dir}) is related to the time derivative of this last term
\[
\frac{dB}{dt}={\bi S}\cdot\left(\frac{d{\bi v}}{dt}\times{\bi v}\right).
\]
This new observable $B$, with dimensions of action, has a positive time derivative for particles and 
a negative time derivative for antiparticles. This sign is clearly related to the sign of $H$.
In the center of mass frame ${\bi P}=0$, $H=\pm mc^2=dB/dt$, with solution $B(t)=B(0)\pm mc^2 t$. 
In units of $\hbar$ this observable represents half the phase of the internal motion
\[
B(t)=B(0)\pm \frac{1}{2}\hbar\omega t.
\]
Because the additional local rotations generated by the $T_i$ commute with the ${\cal P}_D$ group,
the above kinematical variables also span a homogeneous space of the whole ${\cal P}_D\otimes SU(2)_T$ group
and, therefore, they represent the kinematical variables of an elementary system which has this new group
as its kinematical group of space-time symmetries. 

\section{Conclusions and Comments}

We have analyzed the space-time symmetry group of a relativistic model of a Dirac particle.
Matter described by this model ($H>0$ states), is left handed while 
antimatter $(H<0)$, is right handed, 
as far as the relative orientation between the spin and the motion of the charge, 
is concerned. For matter, once the spin direction is fixed, the motion of the charge is counterclockwise
when looking along the spin direction. It is contained in a plane orthogonal to the spin direction,
with the usual sign convention for multivectors in the geometric algebra. 
The motion is clockwise for antimatter.  

This particle has as symmetry group of the Lagrangian ${\cal P}_D\otimes SO(3)_T$ and 
${\cal P}_D\otimes SU(2)_T$ in its quantum description, 
which is larger than the Poincar\'e
group we started with as the initial kinematical group of the model. It contains in its quantum
description, in addition to the Poincar\'e
transformations, a $U(1)$ group which is a unitary representation of the space-time dilations and also 
a $SU(2)_T$ group which is the unitary representation of the symmetry group of 
local rotations of the body frame. The whole group has two Casimir operators
$S^2$, the Casimir of ${\cal P}_D$ and $T^2$ the Casimir of $SU(2)_T$, 
which take the eigenvalues $s=t=1/2$ for the model considered here.

Some of the features we get
have a certain resemblance to the standard model of elementary particles, as far as kinematics is concerned.
If we interpret the generators $T_i$ of the unitary representation of the local rotations 
as describing isospin and the angular momentum
operators ${\bi Z}$ related to the zitterbewegung as describing colour, 
an elementary particle described by this formalism 
is a massive system of spin $1/2$, 
isospin $1/2$, of undetermined mass and charge. It can be in a $s_3=\pm1/2$ spin state and also in a $t_3=\pm1/2$
isospin state.
There are two nonequivalent irreducible representations according to the value of the 
zitterbewegung part of the spin $z$.
It can only be a colourless particle $z=0$ (lepton?) or a 
coloured one $z=1$ in any of three possible colour states $z_3=1,0,-1$, (quark?) but no greater $z$ value
is allowed. 
The basic states can thus also be taken as eigenvectors of ${\bi Z}^2$ but not of $Z_3$, so that
the corresponding colour is unobservable.
There are no possibility of transitions between the coloured and colourless particles because
of the orthogonality of the corresponding irreducible representations. 

Because the eigenvalues of $Z_3$ are unobservable we also have an additional unitary group
of transformations $SU(3)$ which transforms the three $Z_3$ eigenvectors $Y_i^j$ of (\ref{esfhar}) among themselves
and which do not change the $z=1$ value of the eigenstates $\Psi_i$. Nevertheless,
the relationship between this
new $SU(3)$ internal group, which is not a space-time symmetry group, 
and ${\cal P}_D\otimes SU(2)_T$ is not as simple as 
a direct product and its analysis is left to a subsequent research.

This formalism is pure kinematical. We have made no mention to any electro\-magnetic, weak or strong interaction
among the different models. So that, if we find this comparison with the standard model a little artificial, 
the mentioned model of Dirac
particle just represents a massive system of spin $1/2$, spin projection on the body frame
$1/2$, of undetermined mass and charge. It can be in a $s_3=\pm1/2$ spin state and also in a $t_3=\pm1/2$
when projected the spin on the body axis.
There are two different models of these Dirac particles 
according to the value of the orbital or zitterbewegung spin, $z=0$ or 
$z=1$, in any of the three possible orbital spin states $z_3=1,0,-1$, which are unobservable,
but no particle of greater $z$ value is allowed. It is the spin, with its twofold structure orbital and rotational, 
the only intrinsic attribute of this Dirac elementary particle.

\ack{ This work has been partially supported by 
Universidad del Pa\'{\i}s Vasco/Euskal Herriko Unibertsitatea grant  9/UPV00172.310-14456/2002.}

\appendix
\section*{Appendix}
\setcounter{section}{1}

Under infinitesimal time and space translations of parameters $\delta\tau$ and $\delta{\bi b}$, respectively, 
the kinematical variables transform as
\[
t'=t+\delta\tau,\quad {\bi r}'={\bi r}+\delta{\bi b},\quad {\bi v}'={\bi v},\quad \balpha'=\balpha,
\]
so that the selfadjoint generators of translations are
\[
H=i\frac{\partial }{\partial t},\quad {\bi P}=-i\nabla,\qquad [H,{\bi P}]=0.
\]
Under an infinitesimal space-time dilation of normal parameter $\delta\lambda$, they transform in the way:
\[
t'=t+t\delta\lambda,\quad {\bi r}'={\bi r}+{\bi r}\delta{\lambda},\quad {\bi v}'={\bi v},\quad \balpha'=\balpha,
\]
so that the generator takes the form:
\[
D=it\frac{\partial }{\partial t}+i{\bi r}\cdot\nabla=tH-{\bi r}\cdot{\bi P},\qquad [D,H]=-iH,\quad [D,P_j]=-i P_j.
\]
To describe orientation we can represent every element of the rotation group by the three-vector $\balpha=\alpha{\bi n}$,
where $\alpha$ is the rotated angle and ${\bi n}$ is a unit vector along the rotation axis. 
This is the normal or canonical parameterization.
Alternatively we can represent
every rotation by the three-vector ${\brho}=\tan(\alpha/2){\bi n}$. In this case, every rotation matrix takes the form:
\[
R(\brho)_{ij}=\frac{1}{1+\rho^2}\left((1-\rho^2)\delta_{ij}+2\rho_i\rho_j+2\epsilon_{ikj}\rho_k\right).
\]
The advantage of this parameterization is that the composition of rotations $R(\brho')=R(\bmu)R(\brho)$ takes the simple form
\[
\brho'=\frac{\bmu+\brho+\bmu\times\brho}{1-\bmu\cdot\brho}.
\]
Under an infinitesimal rotation of parameter $\delta{\bmu}=\delta\balpha/2$, in terms of the normal parameter, 
the kinematical variables transform:
\begin{eqnarray*}
\delta t&=&0\\
\delta r_{i}&=&-2\epsilon_{ijk}r_j\delta\mu_{k}\\
\delta v_{i}&=&-2\epsilon_{ijk}v_j\delta\mu_{k}\\
\delta\rho_{i}&=&\left(\delta_{ik}+\rho_i\rho_k+\epsilon_{ikl}\rho_l\right)\delta\mu_k,
 \end{eqnarray*}
so that the variation of the kinematical variables per unit of normal rotation parameter $\delta\alpha_k$, is
\begin{eqnarray*}
\delta t_k&=&0\\
\delta r_{ik}&=&-\epsilon_{ijk}r_j\\
\delta v_{ik}&=&-\epsilon_{ijk}v_j\\
\delta\rho_{ik}&=&\frac{1}{2}\left[\delta_{ik}+\rho_i\rho_k+\epsilon_{ikl}\rho_l\right],
 \end{eqnarray*}
and the self-adjoint generators $J_k$, are
\[
J_k=i\epsilon_{ijk}r_j\frac{\partial }{\partial r_i}+i\epsilon_{ijk}v_j\frac{\partial }{\partial v_i}+
\frac{1}{2i}\left(\frac{\partial }{\partial \rho_k}+\rho_k\rho_i\frac{\partial }{\partial \rho_i}+
\epsilon_{ikl}\rho_l\frac{\partial }{\partial \rho_i}\right).
\]
They can be separated into three parts, according to the differential operators involved, with respect to the 
three kinds of kinematical variables ${\bi r}$, ${\bi v}$ and ${\brho}$, respectively:
\[
{\bi J}={\bi L}+{\bi Z}+{\bi W},\]
\[
L_k=i\epsilon_{ijk}r_j\frac{\partial }{\partial r_i},\]
 \begin{equation}
 Z_k=i\epsilon_{ijk}v_j\frac{\partial }{\partial v_i},\quad
W_k=\frac{1}{2i}\left(\frac{\partial }{\partial \rho_k}+\rho_k\rho_i\frac{\partial }{\partial \rho_i}+
\epsilon_{ikl}\rho_l\frac{\partial }{\partial \rho_i}\right).
 \label{eq:Yk}
 \end{equation}
They satisfy the angular momentum commutation rules and commute among themselves:
\[
[L_j, L_k]=i\epsilon_{jkl} L_l,\quad [Z_j, Z_k]=i\epsilon_{jkl} Z_l,\quad[W_j, W_k]=i\epsilon_{jkl} W_l,\]
\[
[{\bi L},{\bi Z}]=[{\bi L},{\bi W}]=[{\bi Z},{\bi W}]=0.
\]
and thus
\[
[J_j, J_k]=i\epsilon_{jkl} J_l,\quad [{\bi J},H]=[{\bi J},D]=0,\quad [J_j, P_k]=i\epsilon_{jkl} P_l.
\]

The above orientation variable $\brho$, under a general boost of velocity ${\bi u}$, transforms as \cite{Rivasbook}
\[
\brho'=\frac{\brho+{\bi F}({\bi u},{\bi v},\rho)}{1+G({\bi u},{\bi v},\rho)}, 
\]
where
\[
{\bi F}({\bi u},{\bi v},\rho)=\frac{\gamma(u)}{1+\gamma(u)}\left({\bi v}\times{\bi u}+{\bi u}({\bi v}\cdot\brho)+
({\bi v}\times\brho)\times{\bi u}\right),
\]
\[
G({\bi u},{\bi v},\rho)=\frac{\gamma(u)}{1+\gamma(u)}\left({\bi v}\cdot{\bi u}+{\bi u}\cdot({\bi v}\times\brho)\right),
\quad \gamma(u)=(1-u^2)^{-1/2}.
\]

Finally, under an infinitesimal boost of value $\delta{\bi u}$, $\gamma(u)\approx1$, the kinematical variables transform:
\begin{eqnarray*}
\delta t&=&{\bi r}\cdot\delta{\bi u}\\
\delta{\bi r}&=&t\delta{\bi u}\\
\delta{\bi v}&=&\delta{\bi u}-{\bi v}({\bi v}\cdot\delta{\bi u})\\
\delta\brho&=&-\left[\brho({\bi v}\cdot\delta{\bi u})+\brho\left(({\bi v}\times{\brho})\cdot\delta{\bi u}\right)
-{\bi v}\times\delta{\bi u}-\delta{\bi u}({\bi v}\cdot{\brho})-\right.\\
&&\left.({\bi v}\times{\brho})\times\delta{\bi u}\right]/2,
 \end{eqnarray*}
and the variation of these variables per unit of infinitesimal velocity parameter $\delta u_j$ is
\begin{eqnarray*}
\delta t_j&=&r_j\\
\delta r_{ij}&=&t\delta_{ij}\\
\delta v_{ij}&=&\delta_{ij}-v_iv_j\\
\delta\rho_{ij}&=&-\frac{1}{2}\left[\rho_jv_i+\rho_i\epsilon_{jkl}v_k\rho_l-\epsilon_{ikj}v_k-\delta_{ij}v_k\rho_k\right],
 \end{eqnarray*}
so that the boost generators $K_j$ have the form
 \begin{eqnarray*}
K_j&=&ir_j\frac{\partial }{\partial t}+it\frac{\partial }{\partial r_j}+i\left(\frac{\partial }{\partial v_j}-v_jv_i\frac{\partial }{\partial v_i}\right)
+\\
&&\frac{1}{2i}\left(\rho_jv_i\frac{\partial }{\partial \rho_i}+\rho_i\epsilon_{jkl}v_k\rho_l\frac{\partial }{\partial \rho_i}-\epsilon_{ikj}v_k\frac{\partial }{\partial \rho_i}
-v_k\rho_k\frac{\partial }{\partial \rho_j}\right)
 \end{eqnarray*}

Similarly, the generators $K_j$ can be decomposed into three parts, 
according to the differential operators involved
and we represent them with the same capital letters as in the case of the ${\bi J}$ operators but with a tilde:
\[
{\bi K}=\widetilde{\bi L}+\widetilde{\bi Z}+\widetilde{\bi W},\quad \widetilde{L}_j=ir_j\frac{\partial }{\partial t}+it\frac{\partial }{\partial r_j},\quad
\widetilde{Z}_j=i\left(\frac{\partial }{\partial v_j}-v_jv_i\frac{\partial }{\partial v_i}\right),
\]
\[
\widetilde{W}_j=\frac{1}{2i}\left(\rho_jv_i\frac{\partial }{\partial \rho_i}+
\rho_i\epsilon_{jkl}v_k\rho_l\frac{\partial }{\partial \rho_i}+\epsilon_{jki}v_k\frac{\partial }{\partial \rho_i}
-v_k\rho_k\frac{\partial }{\partial \rho_j}\right)
\]
They satisfy the commutation rules:
\[
[\widetilde{L}_j,\widetilde{L}_k]=-i\epsilon_{jkl}L_l,\quad [\widetilde{Z}_j,\widetilde{Z}_k]=-i\epsilon_{jkl}{Z_l},\quad
[\widetilde{\bi L},\widetilde{\bi Z}]=[\widetilde{\bi L},\widetilde{\bi W}]=0,
\]
and also
\[
[{K_j},{K_k}]=-i\epsilon_{jkl}{J_l}.
\]

We can check that
\[
\widetilde{\bi Z}={\bi v}\times{\bi Z},\quad
\widetilde{\bi W}={\bi v}\times{\bi W}.
\]

If we define the spin operator ${\bi S}={\bi Z}+{\bi W}$, and the part of the kinematical momentum
$\widetilde{\bi S}=\widetilde{\bi Z}+\widetilde{\bi W}={\bi v}\times{\bi S}$, they satisfy:
\[
[S_j, S_k]=i\epsilon_{jkl} S_l,\quad 
[S_j, \widetilde{S}_k]=i\epsilon_{jkl} \widetilde{S}_l,\quad 
[\widetilde{S}_j,\widetilde{S}_k]=-i\epsilon_{jkl}S_l,
\]
where in the last expression we have used the constraint $v^2=1$. They generate the Lie algebra of a Lorentz group
which commutes with space-time translations $[{\bi S},p^\mu]=[\widetilde{\bi S},p^\mu]=0$.

In the $\brho$ parameterization of the rotation group, the unit vectors of the body frame ${\bi e}_i$, $i=1,2,3$
have the following components:
\[
({\bi e}_i)_j=R(\brho)_{ji},
\]
so that the $T_k={\bi e}_k\cdot{\bi W}$ operators of projecting the rotational angular momentum ${\bi W}$ 
onto the body frame, are given by
 \begin{equation}
T_k=\frac{1}{2i}\left(\frac{\partial }{\partial \rho_k}+\rho_k\rho_i\frac{\partial }{\partial \rho_i}-
\epsilon_{ikl}\rho_l\frac{\partial }{\partial \rho_i}\right).
 \label{eq:Tk}
 \end{equation}
They differ from the $W_k$ in (\ref{eq:Yk}) by the change of $\brho$ by $-\brho$, followed by a global change of sign.
They satisfy the commutation relations
 \begin{equation}
[T_j, T_k]=-i\epsilon_{jkl} T_l.
 \label{eq:Tcom}
 \end{equation}
The minus sign on the right hand side of (\ref{eq:Tcom}) corresponds to 
the difference between the active and passive point of view of transformations. 
The rotation of the laboratory axis (passive rotation) has as generators the ${\bi J}$, which satisfy
$[J_j,J_k]=i\epsilon_{jkl}J_l$. 
The $T_i$ correspond to the generators of rotations of the particle axis (active rotation),
so that, the generators $-T_i$ will also be passive generators of rotations and satisfy $[-T_j,-T_k]=i\epsilon_{jkl}(-T_l)$.

In the normal parameterization of rotations $\balpha=\alpha{\bi n}$, 
if we describe the unit vector ${\bi n}$ along the rotation axis
by the usual polar and azimuthal angles $\theta$ and $\phi$, respectively, 
so that ${\bi n}\equiv(\sin\theta\cos\phi,\sin\theta\sin\phi,\cos\theta)$,
the above $W_i$ generators take the form \cite{RivasS}:
 \begin{eqnarray}
W_1&=&{1\over 2i}\left[2\sin\theta\,\cos\phi\,{\partial\over\partial\alpha}+ 
\left({\cos\theta\,\cos\phi\over\tan(\alpha/2)}-\sin\phi\right){\partial\over\partial\theta}\,-\right.\nonumber\\
&&\left.\left({\sin\phi\over\tan(\alpha/2)\sin\theta}+{\cos\theta\, 
\cos\phi\over\sin\theta}\right)\,{\partial\over\partial\phi}\right],
 \label{eq:Y1}
 \end{eqnarray}
 \begin{eqnarray}
W_2&=&{1\over 2i}\left[2\sin\theta\,\sin\phi\,{\partial\over\partial\alpha}+ 
\left({\cos\theta\,\sin\phi\over\tan(\alpha/2)}+\cos\phi\right){\partial\over\partial\theta}\,-\right.\nonumber\\
&&\left.\left({\cos\theta\,\sin\phi\over\sin\theta}-{\cos\phi\over\tan(\alpha/2)\sin\theta}\right) 
{\partial\over\partial\phi}\right],
 \label{eq:Y2}
 \end{eqnarray}
 \begin{eqnarray}
W_3={1\over 2i}\left[2\cos\theta\,{\partial\over\partial\alpha}-
{\sin\theta\over\tan(\alpha/2)}{\partial\over\partial\theta}+ 
{\partial\over\partial\phi}\right],
 \label{eq:Y3}
 \end{eqnarray}
 \begin{eqnarray}
W^2&=&-\left[{\partial^2\over\partial\alpha^2}+{1\over\tan(\alpha/2)}
{\partial\over\partial\alpha}+\nonumber\right.\\
&&\left.{1\over 4\sin^2(\alpha/2)}\left\{{\partial^2\over\partial\theta^2}+ 
{\cos\theta\over\sin\theta}{\partial\over\partial\theta} 
+{1\over\sin^2\theta}{\partial^2\over\partial\phi^2}\right\}\right],
 \end{eqnarray}
 \begin{eqnarray}
W_+=W_1+iW_2&=&{e^{i\phi}\over 2i}\left[2\sin\theta\,{\partial\over\partial\alpha}+ 
{\cos\theta\over\tan(\alpha/2)}\,{\partial\over\partial\theta} 
+i\,{\partial\over\partial\theta} -
{\cos\theta\over\sin\theta}\,{\partial\over\partial\phi}\;+\nonumber\right.\\
&&\left.{i\over 
\tan((\alpha/2))\sin\theta}\,{\partial\over\partial\phi}\right],
 \end{eqnarray}
 \begin{eqnarray}
W_-=W_1-iW_2&=&{e^{-i\phi}\over 2i}\left[2\sin\theta\,{\partial\over\partial\alpha}+ 
{\cos\theta\over\tan(\alpha/2)}\,{\partial\over\partial\theta} -
i\,{\partial\over\partial\theta} -
{\cos\theta\over\sin\theta}\,{\partial\over\partial\phi}\;-\nonumber\right.\\
&&\left.{i\over 
\tan(\alpha/2)\sin\theta}\,{\partial\over\partial\phi}\right],
 \end{eqnarray}
and the passive $T_i$ generators the form
 \begin{eqnarray}
T_1&=&{-i\over 2}\left[2\sin\theta\,\cos\phi\,{\partial\over\partial\alpha}+ 
\left({\cos\theta\,\cos\phi\over\tan(\alpha/2)}+\sin\phi\right){\partial\over\partial\theta}\;-\nonumber\right.\\
&&\left.\left({\sin\phi\over\tan(\alpha/2)\sin\theta}-{\cos\theta\, 
\cos\phi\over\sin\theta}\right)\,{\partial\over\partial\phi}\right],
 \label{eq:T1}
 \end{eqnarray}
 \begin{eqnarray}
T_2&=&{-i\over 2}\left[2\sin\theta\,\sin\phi\,{\partial\over\partial\alpha}+ 
\left({\cos\theta\,\sin\phi\over\tan(\alpha/2)}-\cos\phi\right){\partial\over\partial\theta}\;-\nonumber\right.\\
&&\left.\left(-{\cos\theta\,\sin\phi\over\sin\theta}-{\cos\phi\over\tan(\alpha/2)\sin\theta}\right) 
{\partial\over\partial\phi}\right],
 \label{eq:T2}
 \end{eqnarray}
 \begin{equation}
T_3={-i\over 2}\left[2\cos\theta\,{\partial\over\partial\alpha}-
{\sin\theta\over\tan(\alpha/2)}{\partial\over\partial\theta}- 
{\partial\over\partial\phi}\right].
 \label{eq:T3}
 \end{equation}  
The $T_i$ are related to the $W_i$ by changing $\alpha$ into $-\alpha$.

The normalised eigenvectors of $W^2=T^2$ and $W_3$ and $T_3$ for $w=t=1/2$, written 
in the form $|w_3,t_3>$, (which are also eigenvectors of $Z^2$ with $z=0$) are written as $|0;s_3,t_3>$
 \begin{eqnarray}
\Phi_1&=&|1/2,-1/2>=i\sqrt{2}\sin(\alpha/2)\sin\theta e^{i\phi},\label{eq:Fi1}\\
\Phi_2&=&|-1/2,-1/2>=\sqrt{2}\left(\cos(\alpha/2)-i\cos\theta\sin(\alpha/2)\right)\\
\Phi_3&=&|1/2,1/2>=-\sqrt{2}\left(\cos(\alpha/2)+i\cos\theta\sin(\alpha/2)\right),\\
\Phi_4&=&|-1/2,1/2>=-i\sqrt{2}\sin(\alpha/2)\sin\theta e^{-i\phi}.\label{eq:Fi4}
 \end{eqnarray}
The rising and lowering operators $W_{\pm}$ and the corresponding $T_{\pm}$ transform them among each other. 
$\{\Phi_1,\Phi_2\}$ are related by the $W_{\pm}$, and similarly the $\{\Phi_3,\Phi_4\}$ while the sets 
$\{\Phi_1,\Phi_3\}$ and $\{\Phi_2,\Phi_4\}$ are separately related by the $T_{\pm}$.
For instance
\[
W_-\Phi_1=\Phi_2,\quad W_-\Phi_2=0,\quad W_-\Phi_3=\Phi_4,\]
\[
 T_-\Phi_1=\Phi_3,\quad T_-\Phi_3=0,\quad T_-\Phi_2=\Phi_4.
\]
They form an orthonormal set with respect to the normalised invariant measure defined on $SU(2)$
\[
dg(\alpha,\theta,\phi)=\frac{1}{4\pi^2}\sin^2(\alpha/2)\sin\theta\,d\alpha\,d\theta\,d\phi,\]
\[
 \alpha\in[0,2\pi],\quad \theta\in[0,\pi],\quad \phi\in[0,2\pi].
\]
\[
\int_{SU(2)} \,dg(\alpha,\theta,\phi)=1.
\]
The wave function $\psi$ can be separated in the form
\[
\psi(t,{\bi r},{\bi v},{\balpha})=\sum_{i=1}^{i=4}\phi_i(t,{\bi r})\chi_i({\bi v},{\balpha})
\]
where the four $\chi_i$ can be classified according to the eigenvalues $|s_3,t_3>$.
The functions $\phi_i(t,{\bi r})$ can be chosen as eigenfunctions of the Klein-Gordon operator \cite{Rivasbook}
\[
p_\mu p^\mu \phi_i(t,{\bi r})=m_i^2\phi_i(t,{\bi r}).
\]
The functions $\chi({\bi v},{\balpha})$ can also be separated because the total spin ${\bi S}$ with $s=1/2$, is the sum
of the two parts ${\bi S}={\bi Z}+{\bi W}$, with $[{\bi Z},{\bi W}]=0$, 
so that since the ${\bi W}$ part contributes with $w=1/2$ then 
the ${\bi Z}$ part contributes with $z=0$ or $z=1$. The $z=0$ contribution corresponds to the 
functions $\chi_i(\balpha)$
independent of the velocity variables and the orthonormal set are the above $\Phi_i$, $i=1,2,3,4$, which can also 
be written in the form $|z;s_3,t_3>$, with $z=0$.

Because ${\bi Z}=-i{\bi v}\times\nabla_v$, for the $z=1$ part the eigenvectors of $Z^2$ and $Z_3$ are
the spherical harmonics $Y^i_1(\widetilde{\theta},\widetilde{\phi})$, $i=-1,0,1$.
The variables $\widetilde{\theta}$ and $\widetilde{\phi}$ represent the direction 
of the velocity vector ${\bi v}$. 
Because $[Z_i,W_j]=0$, we can again separate the variables in the functions $\chi({\bi v},\balpha)$.
In this case the $\chi({\bi v},\balpha)=\sum \phi_i(\widetilde{\theta},\widetilde{\phi})\lambda_i(\alpha,\theta,\phi)$.
The four orthonormal vectors, eigenvectors of $S_3$, $Z^2$ with $z=1$ and $T_3$, $|1;s_3,t_3>$, are now
 \begin{eqnarray}
{\Psi}_1&=&|1;1/2,1/2>=\frac{1}{\sqrt{3}}\left(Y_1^0(\widetilde{\theta},\widetilde{\phi})\Phi_1-\sqrt{2}Y_1^1(\widetilde{\theta},\widetilde{\phi})\Phi_2\right),\label{eq:Fit1}\\
{\Psi}_2&=&|1;-1/2,1/2>=\frac{1}{\sqrt{3}}\left(-Y_1^0(\widetilde{\theta},\widetilde{\phi})\Phi_2+\sqrt{2}Y_1^{-1}(\widetilde{\theta},\widetilde{\phi})\Phi_1\right),\\
{\Psi}_3&=&|1;1/2,-1/2>=\frac{1}{\sqrt{3}}\left(Y_1^0(\widetilde{\theta},\widetilde{\phi})\Phi_3-\sqrt{2}Y_1^1(\widetilde{\theta},\widetilde{\phi})\Phi_4\right),\\
{\Psi}_4&=&|1;-1/2,-1/2>=\frac{1}{\sqrt{3}}\left(-Y_1^0(\widetilde{\theta},\widetilde{\phi})\Phi_4+\sqrt{2}Y_1^{-1}(\widetilde{\theta},\widetilde{\phi})\Phi_3\right).\label{eq:Fit4}
 \end{eqnarray}
where the $\Phi_i$ are the same as the ones in (\ref{eq:Fi1}-\ref{eq:Fi4}) and the spherical harmonics $Y^i_1(\widetilde{\theta},\widetilde{\phi})$ are
\begin{equation}
Y_1^1=-\sqrt{\frac{3}{8\pi}}\sin(\widetilde{\theta})e^{i\widetilde{\phi}},\;\;
Y_1^0=\sqrt{\frac{3}{4\pi}}\cos(\widetilde{\theta}),\;\;
Y_1^{-1}=\sqrt{\frac{3}{8\pi}}\sin(\widetilde{\theta})e^{-i\widetilde{\phi}}.
\label{esfhar}
\end{equation}
The $Z_i$ operators are given by
\[
Z_1=i\sin{\widetilde\phi}\frac{\partial }{\partial\widetilde\theta}+i\frac{\cos{\widetilde\theta}}{\sin{\widetilde\theta}}\cos{\widetilde\phi}\frac{\partial }{\partial\widetilde\phi},\quad
Z_2=-i\cos{\widetilde\phi}\frac{\partial }{\partial\widetilde\theta}+i\frac{\cos{\widetilde\theta}}{\sin{\widetilde\theta}}\sin{\widetilde\phi}\frac{\partial }{\partial\widetilde\phi},\]
\[
Z_3=-i\frac{\partial }{\partial\widetilde\phi}.
\]
The raising and lowering operators $Z_{\pm}$ are
\[
Z_{\pm}=e^{\pm i\widetilde{\phi}}\left(\pm\frac{\partial }{\partial \widetilde{\theta}}+i\frac{\cos\widetilde{\theta}}{\sin\widetilde{\theta}}\frac{\partial }{\partial\widetilde{\phi}}\right),
\]
so that
\[
Z_-Y_1^1=\sqrt{2}\;Y_1^0,\quad Z_-Y_1^0={\sqrt{2}}\;Y_1^{-1}.
\]
The four spinors ${\Psi}_i$ are orthonormal with respect to the invariant measure
\[
dg(\alpha,\theta,\phi\,;\widetilde{\theta},\widetilde{\phi})=\frac{1}{4\pi^2}\sin^2(\alpha/2)\sin\theta\sin\widetilde\theta\,d\alpha\,d\theta\,d\phi\,d\widetilde{\theta}d\widetilde{\phi}\]
\[
 \alpha\in[0,2\pi],\quad \widetilde{\theta},\theta\in[0,\pi],\quad \widetilde{\phi},\phi\in[0,2\pi].
\]
Similarly as before, the rising and lowering operators $S_{\pm}=Z_{\pm}+W_{\pm}$ and the corresponding $T_{\pm}$ 
transform the ${\Psi}_i$ among each other. 
In particular the $\{{\Psi}_1,{\Psi}_2\}$ are related by the $S_{\pm}$, and similarly 
the $\{{\Psi}_3,{\Psi}_4\}$ while the sets 
$\{{\Psi}_1,{\Psi}_3\}$ and $\{{\Psi}_2,{\Psi}_4\}$ are separately related by the $T_{\pm}$. This is the reason
why the general spinor in this representation is a four-component object.

In the $z=0$ basis $\Phi_i$ (\ref{eq:Fi1}-\ref{eq:Fi4}), the spin operators and the basis vectors of the body frame 
take the form:
\[
{\bi S}=\frac{1}{2}\pmatrix{\bsigma&0\cr 0&\bsigma}={\bi W},
\]
\[
T_1=\frac{1}{2}\pmatrix{0&\ID\cr \ID&0},\quad
T_2=\frac{1}{2}\pmatrix{0&-i\ID\cr i\ID&0},\quad
T_3=\frac{1}{2}\pmatrix{\ID&0\cr 0&-\ID},
\]
\[
{\bi e}_1=\frac{-1}{3}\pmatrix{0&\bsigma\cr \bsigma&0},\quad
{\bi e}_2=\frac{-1}{3}\pmatrix{0&-i\bsigma\cr i\bsigma&0},\quad
{\bi e}_3=\frac{-1}{3}\pmatrix{\bsigma&0\cr 0&-\bsigma},
\]
in terms of the Pauli $\bsigma$ matrices and the $2\times2$ unit matrix $\ID$.

In the $z=1$ basis ${\Psi}_i$ (\ref{eq:Fit1}-\ref{eq:Fit4}), the operators $S_i$ and $T_i$ take 
the same matrix form as above, while the
${\bi e}_i$ are
\[
{\bi e}_1=\frac{1}{9}\pmatrix{0&\bsigma\cr \bsigma&0},\quad
{\bi e}_2=\frac{1}{9}\pmatrix{0&-i\bsigma\cr i\bsigma&0},\quad
{\bi e}_3=\frac{1}{9}\pmatrix{\bsigma&0\cr 0&-\bsigma}.
\]
In all cases, the 6 hermitian traceless matrices $S_i$, $T_j$, the nine hermitian 
traceless matrices ${e_i}_j$ and the $4\times4$ unit
matrix are linearly independent and they completely define a hermitian basis 
for Dirac's algebra, so that any other translation
invariant observable of the particle will be expressed as a real linear combination 
of the above 16 hermitian matrices. We used this fact in reference \cite{Rivas2}
to explicitely obtain Dirac's equation for this model.

Both representations are orthogonal to each other, $<\Phi_i|\Psi_j>=0$, 
and they produce two different irreducible representations of the group, 
so that they describe two different
kinds of particles of the same spin 1/2.

The matrix representation of the $Z_i$ and $W_i$ operators in the basis $\Psi_i$ are given by
\[
{\bi Z}=\frac{2}{3}\pmatrix{\bsigma&0\cr 0&\bsigma},\quad {\bi W}=\frac{-1}{6}\pmatrix{\bsigma&0\cr 0&\bsigma},
\]
although the $\Psi_i$ are not eigenvectors of $Z_3$ and $W_3$.



\section*{References}
\begin{thebibliography}{10}
 \bibitem{Rivasbook} Rivas M 2001 {\em Kinematical theory of spinning particles}, 
(Dordrecht: Kluwer).\\ See also the Lecture notes of the course {\it Kinematical formalism of elementary
spinning particles} delivered at JINR, Dubna, 19-23 September 2005. ({\it Preprint} physics/0509131)

 \bibitem{Rivas2} Rivas M 1994 {\it J. Math. Phys.} {\bf 35} 3380.

\bibitem{dyn}  Rivas M 2003 {\it J. Phys. A: Math. Gen.} {\bf 36} 4703 ({\it Preprint} physics/0112005).

 \bibitem{RivasS} See reference \cite{Rivasbook} section 4.3.


\bibitem{Abellanas} Abellanas L and Martinez Alonso L 1975 {\it J. Math. Phys.} {\bf 16} 1580.
\end{thebibliography}
\end{document}